\documentclass[letter,12pt]{article}
\usepackage{graphicx,amssymb,amsmath,epstopdf,color,hyperref,tocloft}

\def\beq{\begin{equation}}
\def\eeq{\end{equation}}
\def\bea{\begin{eqnarray}}
\def\eea{\end{eqnarray}}

\def\gev{\, {\rm GeV}}
\def\tev{\, {\rm TeV}}
\newcommand{\gsim}{\lower.7ex\hbox{$\;\stackrel{\textstyle>}{\sim}\;$}}
\newcommand{\lsim}{\lower.7ex\hbox{$\;\stackrel{\textstyle<}{\sim}\;$}}

\def\be{\begin{equation}}
\def\ee{\end{equation}}
\def\bea{\begin{eqnarray}}
\def\eea{\end{eqnarray}}

\def\FTone#1{{\rm FT}[#1]}
\def\FTtwo#1#2{{\rm FT}[#1\, |\, #2]}

\begin{document}

\begin{flushright}
July 5, 2021
\end{flushright}

\vspace{0.07in}

\noindent
\begin{center}

{\bf\Large Evaluation and Utility of Wilsonian Naturalness}

\vspace{0.5cm}
{ James D. Wells}

{\it Leinweber Center for Theoretical Physics \\
University of Michigan, Ann Arbor, MI 48109 USA} \\

\end{center}

\vspace{0.5cm}

\noindent
{\it Abstract:} 
We demonstrate that many  Naturalness tests of particle theories discussed in the literature can be reformulated as straightforward algorithmic finetuning assessments  in the matching of Wilsonian effective  theories above and below particle mass thresholds.  Implications of this EFT formulation of Wilsonian Naturalness are discussed for several theories, including the Standard Model, heavy singlet scalar theory, supersymmetry, Grand Unified Theories, twin Higgs theories, and theories of extra dimensions. 
We argue that the Wilsonian Naturalness algorithm presented here constitutes an unambiguous, {\it a priori}, and meaningful test that the Standard Model passes and which ``the next good theory" of particle physics is very likely to pass.

\bigskip
\tableofcontents

\vfill\eject

\section{Naturalness and theory viability}

There are an infinite number of theories that are perfectly compatible with all known data, with the Standard Model (SM) being only one version of that, or rather one class of such theories. This is certainly true for all high-energy collider physics data, which is a key observables target for particle physics theories, as the decoupling nature of viable beyond the Standard Model (BSM) theories allows their preservation. Just push up the ``new physics" scales of these BSM theories and the theories are safe from experimental incompatibility. This is the case for supersymmetry, composite Higgs theories, extra dimensional theories, grand unified theories, and many more. 

This underdetermination problem is of course well-known by researchers. Attempts to make a rank order among good theories by invoking such tools as  Ockham's razor, parsimoniousness, ``beauty", consilience, etc.\ are largely unsatisfying to a rigorous mind. Nevertheless, we will try to do exactly that here, but perhaps with a half-rigorous mind by invoking some probability arguments against  coincidences among random numbers. The set of ideas along these lines goes under the name of ``Naturalness criteria" (see, e.g.,~\cite{Giudice:2008bi,deGouvea:2014xba}), which as we will review below is intimately connected to finetuning.

 Criteria for passing a Naturalness test have been given in the past, even though little discussion has been given to what exactly are the implications for a theory when it does not pass a Naturalness test. The case for a meaningful connection between finetuning assessments, the engine under the hood of Naturalness, and probability has been made in ref.~\cite{Wells:2018yyb}, which  connected low probability with high finetuning of $Z$ with respect to large $X$ and $Y$ in $Z=X-Y$. This is generically the approximate form that gives rise to all finetunings across EFT boundaries. However, there it was stated that a critical feature to legitimacy is the {\it a priori} definition of any finetuning test used in the assessment of Naturalness. In other words, the test needs to be in place before the theory, or one risks manipulating the Naturalness test to serve propaganda purposes.

In this paper we define a general, algorithmic, and {\it a priori} test for Naturalness that serves all the requirements of~\cite{Wells:2018yyb}. The test is based on computing finetunings across particle mass thresholds of a theory whose expression (fields and operators) and calculational scheme is presented according to an {\it a priori} decided-upon convention. We find that once we do that, many of the Naturalness criteria employed in the literature that looked disparate and even {\it ad hoc} are very nearly exactly what one finds when employing what we will call the Wilsonian Naturalness algorithm. If there are such possibilities to make good a would-be Unnatural theory, then this would be another example of a new theory principle recognized to solve the Wilsonian Naturalness problem, thus reenforcing the value of articulating Wilsonian Naturalness assessments. The Naturalness tests that come automatically under this larger Wilsonian Naturalness algorithm framework include finetuning of $m_Z^2$ in supersymmetric theories, doublet-triplet splitting problem in Grand Unified theories, and quadratic sensitivity problem to a heavy scalar singlet. We even automatically get similar conclusions about the would-be Naturalness failure of the SM if the Higgs mass were much lighter (lighter than a few GeV), reminiscent of  the ``finite Naturalness" claims~\cite{Farina:2013mla}.

In the following sections we will first describe what is meant by a theory here, since the Naturalness status is a condition on the theory itself and not directly on observables, so a clear understanding of what constitutes a theory is a required first step. We then describe an algorithmic way to compute all the threshold finetunings in a particle physics theory. After that we connect threshold finetunings, and levels of finetuning, to the notion of Wilsonian Naturalness. We then apply these considerations to several elementary particle theories: Standard Model, supersymmetry, grand unified theories, added scalar singlet, and theories of extra dimensions. A chief goal is to determine in what circumstances we believe that the Wilsonian Naturalness test  may relegate these theories to a lower status, i.e., unlikely to be a good description nature. The final conclusion section summarizes the findings.

\section{Indexed theories of a theory class}

The question of whether a theory is Natural is a question internal to the theory under consideration. For that reason we need to specify what is meant by a theory in order to articulate this internal analysis. 

For us, a theory $T$ is an algorithm that maps a specified domain of state descriptors $\{\xi_i\}$ to a collection of target observables $\{ {\cal O}_k\}$:
\beq
T: \{\xi_i\}\longrightarrow \{ {\cal O}_k\}
\eeq
By state descriptors we mean contingent variables for the state of the system. One could call it the in-state of the system $\psi_{\rm in}$. For example, an electron with momentum $p_1$ and a positron with momentum $p_2$ are the state descriptors, which are known and interpretable by the theory. The output target observables are, for example, total cross-section into $\mu^+\mu^-$, forward-backward asymmetries, differential angular cross-section of $e^+e^-$, and many more.

In our language the Standard Model (SM) is not a single theory but a well-defined class of theories. The reason is that in order to do a calculation a set of theory parameters must be set $\vec g=\{ g',g,y_t,\lambda,\theta_{\rm QCD},\cdots\}$. Thus, the class of theories we call the SM becomes a theory when the $\vec g$ are set to specific values. In this sense, the theory parameters index the infinite number of SM theories: $T_{\vec g}={\rm SM}(\vec g)$.
One can then find the parameter space of $\vec g$ values that are in accord with experiment. 

We therefore can speak in general of a class $T$ of theories $T_{\vec g}$ indexed by $\vec g$ with a parameter space of $\vec g$ that is consistent with experiment. For any given choice of parameters $\vec g$ that has been determined to be in accord with experiment, one then can perform an internal analysis on the theory to ask if it satisfies a quality or suffers a defect that we have argued to be important to identify.

In our case we will do an analysis of theories $T_{\vec g}$ to try to discover whether it passes the Wilsonian Naturalness test. The answer to this question may appear then to be completely divorced from observables and data, but it is not. The easiest way to understand how experiment is crucial in all of these considerations is to realize that it is quite possible that there is a parameters space domain of $\vec g$, which we can call  $\vec g_A$, that is inconsistent with experiment yet internal analysis says that $T_{\vec g_A}$ passes the Wilsonian Naturalness test. On the other hand, for the entire domain of $\vec g$ that is consistent with experiment, internal analysis says that those $T_{\vec g}$ theories fail the Wilsonian Naturalness test. Therefore, it is the combination of experimental data (observables) and internal analysis (Wilsonian Naturalness test) that calls into question the entire class of theories $T_{\vec g}$. Neither one of them alone could do it. In other words, Naturalness claims and its impact on the viability of a full class of theories is just as much about experimental data as it is about internal analysis of the theory.


\section{Threshold finetuning and Naturalness tests}

In this section we first define what is meant by threshold finetuning in a theory and give an algorithm for computing it. We then use the numerical output of the threshold finetuning measure to argue that some theories fail the Wilsonian Naturalness test and should be dismissed. We do this in the context of Wilsonian effective theories~\cite{Burgess:2021}.

By threshold we mean the mass scale of a particle $\Phi_H$ in the spectrum of the theory. In QFT we can define a high-scale theory with lagrangian ${\cal L}_H$ to be one where the particle is a freely propagating degree of freedom. This theory is valid for momentum scales above and below the threshold particle's mass; however, as is well known in QFT, computations of observables for energies well below a particle's mass are beset with large logarithms that are calculationally difficult to handle. However, from the Appelquist-Carrazone decoupling theorem we know that the effects of the heavy state should decouple at low energies, and one can write a new low-scale theory with lagrangian ${\cal L}_L$ that does not have that state. The lagrangian couplings of the high scale theory, such as Yukawa couplings, gauge couplings, and mass terms, we denote collectively by $\vec g_H$, and the lagrangian couplings of the low scale theory by $\vec g_L$. Let's call the heavy state $\Phi_H$ and all other light states $\phi_i$. What we have then is
\bea
{\cal L}_H(\Phi_H,\varphi_i; \vec g_H) & & {\rm high~scale~theory~with}~\Phi_H, {\rm ~and}\\
{\cal L}_L(\varphi_i; \vec g_L)& & {\rm low~scale~theory~without~}\Phi_H~{\rm (integrated~out)}
\eea
where the $\Phi_H$ field has been integrated out. Note, no coupling $\vec g_L$ in the low-scale theory is the ``same" as any coupling $\vec g_H$ in the high-scale theory despite appearances. There are schemes in which they are the same at the threshold, but in principle they are different couplings with different renormalization group flows, etc.

Let us define the mass of $\Phi_H$ to be $M$, which sets our definition of the threshold scale at the boundary between ${\cal L}_H$ and ${\cal L}_L$. Upon integrating out the $\Phi_H$ state at the scale $M$ one finds matching conditions between the $\vec g_L$ couplings and $\vec g_H$ couplings at that scale. Any given low-scale theory coupling $g_{Li}$ is matched to a function of high-scale theory parameters at the scale $M$:
\beq
g_{Li}(M)=F_i(\vec g_H(M))
\eeq
where the $F_i$ functions are to be determined by computation. 

At this point to make notation simpler we drop the argument of the couplings and define $g_{Li}\equiv g_{Li}(M)$ and $g_{Hj}\equiv g_{Hj}(M)$, and we define a series of threshold finetunings~\cite{Barbieri:1987fn} associated with every low-scale theory parameter:
\beq
\FTtwo{g_{Li}}{g_{Hj}}= \left| \frac{g_{Hj}}{g_{Li}}\frac{\partial g_{Li}}{\partial g_{Hj}}\right|_{\mu^2= M^2}
\eeq
We can define the finetuning associated with low-energy parameter $g_{Li}$ to be
\beq
\FTone{g_{Li}}={\rm max}_k~\FTtwo{g_{Li}}{g_{Hk}}
\eeq
where $\mu^2$ is the renormalization group scale at which the parameters and logarithms are evaluated.
This equation looks rather complex with many calculations to do, but we will have  one primary focus and that is the Higgs boson mass\footnote{The cosmological constant is another coefficient of worry for finetuning and naturalness~\cite{Burgess:2013ara}. However, it is a dimension-zero operator and deeply connected to issues of quantum gravity, which are beyond our scope.  For this reason, we do not consider it further when assessing Naturalness within our particle physics theories.}. To be precise we mean the $m^2$ coupling of the $H^\dagger H$ operator of the Standard Model,
\beq
{\cal L}_{\rm SM}=-m^2H^\dagger H+\cdots.
\eeq
  Let us call the finetuning across some heavy threshold $M$ this simply as ``Higgs finetuning" defined as
\beq
{\rm Higgs~finetuning}~\equiv \FTone{m^2}={\rm max}_k \left| \frac{g_{Hk}}{m^2}\frac{\partial m^2}{\partial g_{Hk}}\right|_{\mu^2=M^2}
\eeq
where the threshold $M$ is the one that gives the largest threshold finetuning in the theory.

For a threshold finetuning definition to be useful in this circumstance it must be defined {\it a priori}, meaning that we cannot find a special scheme for writing the lagrangian, or a special scheme for renormalization for the sole purpose of trying to minimize the finetuning. Instead, we suggest merely using the most standard approach to calculational scheme, $\overline{MS}$ and $\overline{DR}$ (if supersymmetric), and an {\it a priori} straightforward/textbook way of writing the lagrangian operators with their coefficients of $g$, $m^2$, $\lambda$, $y_t$, $\theta_{\rm QCD}$, etc., and simply apply the above definitions to find the finetunings across each threshold as one flows from the UV to the IR\footnote{It must be emphasized that the {\it arbitrariness} of the chosen elements for the algorithm is a feature, as it samples on the (mental) distribution of an infinite set of other possibilities for how a theory might be constructed or how a calculational procedure might be implemented. Nevertheless, maintaining {\it consistency} of the pre-defined definition is also key so as to not let creep in case-by-case re-assessments of a finetuned theory.}. What we present here is an algorithm for Wilsonian Naturalness and a theory under that calculation has an unambiguous, fixed status, independent of how important one might think that status is, which is to be addressed next. 

The key claim here on the utility of this algorithm is that it is highly unexpected that the finetunings across the thresholds should ever be greater than some very large value, such as $10^4$ or so. The intuition for this is set by probability calculations based on flat priors of lagrangian parameters~\cite{Wells:2018yyb}. As explained in~\cite{Wells:2018yyb}, as we consider higher and higher finetunings the assumed distributions of the lagrangian parameters become more and more irrelevant to our ability to declare that the relationship between parameters that gave rise to such a large finetuning is unlikely. 

An important question is what does one mean by a distribution over parameters. First, as mentioned above, the existence of a reasonably smooth distribution of possible choices of the theory parameters is all that is needed. We do not need to know the precise form of the distribution. Furthermore, the notion of a distribution here is equivalent to calling the parameters contingent. The parameters could be contingent in several senses. In the first sense, we say that they are selected on a landscape of possibilities, which is motivated by string theory considerations of low-energy parameters being one of an extraordinarily large but finite number of sets of possibilities~\cite{Hebecker:2021}. If work related to anthropic principle has no other value it has certainly shown that we can have qualitatively the same universe and features for different choices of the parameters (i.e., different electron mass, different force couplings, and so on), even if we assume the static properties of the universe are rigidly set (existence of electrons, photons, $SU(3)$ gauge theory, etc.). Thus, there is no incompatibility in conceiving of other choices, and no incompatibility in calling the parameters of the theory contingent\footnote{A theory itself may be contingent, just as its parameters are. There may be an infinite number of theories that equally well describe all of particle physics as we know it but based on very different mathematics and constructions, such that the parameters across those theories have a distribution. In that case, fluctuations in our brains selected out the current SM construction that we learn in school, and the SM parameters would be merely numbers selected on a distribution of theories without any need to invoke multiverse distributions.}, and no incompatibility in conjecturing that they have been selected by some deeper principle or mechanism (landscape solutions, baby universes, multiverse, Zeus, etc.). 

Although we believe that it can be justified reasonably that high finetuning signifies that a conjectured theory is unlikely to be fruitful, the real efficacy of finetuning is whether or not it works to separate wheat theories from chaff theories. With the definition given above, we are able to answer that question not only in the future (will any of our highly finetuned conjectured theories pan out?) but also in the present (do any of our fruitful theories, such as the SM, have a high threshold finetuning problem?). 

In the case of the SM there are a dozens of nontrivial tests of our conjecture that threshold finetuning should remain low. One can compute the threshold finetunings of every low-energy parameter across the many thresholds starting with $m_{\rm top}$, and continuing down to $m_h$, $m_Z$, $m_W$, $m_b$, and so on. As we will show in the section devoted to the SM example, the SM passes  those tests and therefore passes the overall Wilsonian Naturalness test. This is exactly what we want from this conjecture that purports to identify dismissable theories. The fruitful and accepted SM theory satisfies a very large number of tests that it could have failed. This is the data that supports the Wilsonian Naturalness test presented here.

Before we describe several examples, let us introduce one additional useful definition to help us quantify finetuning. A theory is ``level-X finetuned" if the maximum finetuning across every threshold of the theory is $10^X$.
\beq
\FTone{m^2}=10^X~~\longrightarrow~~{\rm Level\mbox{-}X~finetuned~theory.}
\eeq
The $X$ value is the numerical datum that comes out of the Wilsonian Naturalness assessment of the theory.
 Using~\cite{Wells:2018yyb} as a guide, I consider level-6 and higher theories to be Unnatural (i.e., fail the Wilsonian Naturalness test) and  should be labeled as such. Level-0,1,2  theories pass the Wilsonian Naturalness test in our view. One can imagine many different views on whether Level-3, Level-4 and Level-5 theories pass the Wilsonian Naturalness test.  It is hard to argue that they are hopelessly improbable at that level and so in the spirit of conservativeness they should not be labeled Unnatural, but they should be looked upon with some suspicion and they likely need other good theory qualities (consilience, additional explanatory power compared to competitors, etc.)\ to remain in the good graces of physicists. The reader may disagree with my assessments for viability of each level, but I hope they would think hard about what level is ok and what is not. One can start by asking if a Level-13 theory of the future is likely, without any extra principle brought in to explain its large finetuning. If you say ``no way is a Level-13 theory going to happen," like I do, then ask yourself about Level-12, and keep going until you get nervous about being too rigidly judgmental about the viability of a hypothesized new theory. For me, that is Level-6, as stated above.
 
 Let us summarize our definition of Wilsonian Naturalness:
 \begin{quote}
 {\bf Endo-Natural theory}: An endo-Natural theory is one where the finetunings are not high (all are, say, level-4 or lower\footnote{Upon surveying the literature of physicists intuitions, any choice of level-3 through level-6 has been deemed a defensible upper bound on tolerable finetuning.}) across all its particle thresholds when matching EFTs above and below the thresholds, according to the {\it a priori} defined algorithms of assessment discussed above. \\ \\
 {\bf Exo-Natural theory}: A theory may have large finetuning across threshold(s), but those finetunings are explained in principle and are not accidental. This case has no implication of low probability despite its large finetuning, and we call the corresponding theory exo-Natural. \\ \\
 {\bf Wilsonian Natural theory}: A Wilsonian Natural theory is one that is either endo-Natural or exo-Natural.
 \end{quote}

 Exo-Natural theories are ones with strong UV/IR correlations embedded in them, which in principle can be discerned. We do not discuss them much in what follows, but let us illustrate here an exo-natural theory by considering the  Arkani-Hamed-Harigaya theory~\cite{Arkani-Hamed:2021xlp}, which was introduced as a possible explanation of the the $g-2$ anomaly~\cite{g2anomaly}. By virtue of finetuned cancellations in the dimension-six operators that give rise to $g-2$, the model ``violates the Wilsonian notion of naturalness" in the words of the authors. In our language,  the violation is not a violation of Wilsonian Naturalness but rather a violation of endo-Naturalness which comes at the matching interface of the second highest mass state $S$, where the EFT has a large coefficient of a $g-2$ inducing dimension-six operator that is nearly exactly cancelled by integrating out the $S$ field at the matching threshold scale $m_S$. But it is this violation of endo-Naturalness that points to the correct supposition that this was no accidental finetuned cancellation and the EFT under consideration (just below the heavier $L$ particle mass) is incomplete and there are surely new particle(s) or new principles at play that {\it are not manifest within the EFT and are yet to be found.} Indeed that is the case. Thus, the theory is an exo-Natural theory, and is therefore Wilsonian Natural. It is for these reasons that it must be emphasized that there is no claim here that endo-Naturalness must always hold in any theory. Rather, the claim is on the implication of when endo-Naturalness does not hold: there are new particle(s) or principle(s) yet to be discovered because the theory is secretly exo-Natural -- i.e., always Wilsonian Natural.

In summary, let us formulate the main conjecture of the paper as follows:
\begin{quote}
{\bf Wilsonian Naturalness conjecture}: Large accidental finetunings in EFT matching across a particle threshold is highly improbable.
Any such finetuning that may occur should be pursued as a sign for the existence of new particle(s) or principle(s) that render the large finetuning as a non-accidental result (i.e., it is secretly an exo-Natural theory). Furthermore, any conjectured theory that relies on large, unexplained accidental finetuning in EFT matching across particle threshold(s) is unlikely to be a good description of nature. Summary: Wilsonian Naturalness is expected to be satisfied by the next useful theory of nature beyond the Standard Model\footnote{Of course, one is free to speculate on the possibility that a more radical idea could come along that does not involve even lagrangians or Wilsonian effective field theories in any currently recognizable way, where Wilsonian Naturalness tests would not be directly applicable.}.
\end{quote}
The subject of the rest of this paper is to illustrate the meaning and value of this conjecture.

\section{The Standard Model}
\label{sec:SM}

Statements abound in the literature that ``the SM suffers from the Naturalness problem", or more or less equivalently, the hierarchy problem and the finetuning problem~\cite{Borrelli:2019}. We need to find a theory that ``cures the SM's naturalness problem" is another common refrain. However, there is no place for such talk when it comes to a highly successful theory like the SM. If some dreamt-up criteria ends up labeling the SM with a Naturalness problem, then I want every other theory I come up with to also have a Naturalness problem just like it.

There is no reasonable sense in which the SM has a naturalness problem or any other problem intrinsic to the theory itself such that it should be relegated to lower status. It is a successful first-class theory, culminating in the discovery of the Higgs boson which was a non-trivial corroboration of one of its peculiar features. However, one could try to expand beyond the  surface meaning of ``Naturalness problem" and say something that is reasonable: The SM has no principle or mechanism suggested within it that can protect the Higgs boson mass if new states that couple to it are introduced at large scales, and since we have no reason to believe that such new states are forbidden in nature,  we realize that there is a problem -- a ``Naturalness problem of proliferation of new states that we generically expect" -- in understanding how such new states could exist while the Higgs boson is so light. We can call this formulation the ``proliferation naturalness problem" and it was argued for in~\cite{Wells:2016luz}. As we will discuss in sec.~\ref{sec:Xdim}, this ``proliferation problem" is what is implicitly behind declaring a theory to have a ``hierarchy problem", where the hierarchy is between the SM states and exotic new states generically expected to exist in nature well above the weak scale.

The above articulation of the ``proliferation naturalness problem" of the SM is really about contemplating simple new theories that are concatenations of the SM, and it by no means suggests that the SM itself is unnatural. The SM has done its job, and whatever Naturalness test one implements the SM better pass it to be a worthwhile test. It is up to any new theory, on the other hand, which might include new states at higher mass scales, to make sure it passes its own Naturalness test.

In what sense can we be sure that the SM passes its Naturalness test? In other words, how does one determine that the criteria set forward earlier when applied to the SM does not condemn it inappropriately as an unnatural theory? The answer is by straightforward calculation. At every particle mass threshold ($m_t$, $m_h$, $m_Z$, $m_W$, $m_b$, etc.)\ compute the low-energy (theory below mass threshold) couplings $\vec g_L$ in terms of the high-energy (theory above mass threshold) couplings $\vec g_H$ and compute the threshold finetunings using $\overline{MS}$ renormalized parameters. The answer is that there are no large threshold finetunings in the theory at all, and it therefore it passes the Naturalness test. The Wilsonian Naturalness algorithm correctly retains the SM.

It should be noted that the SM had very many opportunities to fail its test, since there was always a possibility that among the many matching conditions at the numerous thresholds one could have been highly finetuned. No large finetuning was found, but it could have been.

Perhaps the best opportunity for the finetuning to have manifest itself is across the top quark threshold. The matching of the $m^2$  above and below the top mass after electroweak symmetry breaking requires us to inspect the $m_h^2$ coefficient of $\frac{m_h^2}{2}h^2$ operator above and below the top mass. One finds the leading term to be 
\bea
m^2_h(m_t)_L & = &m^2_h(m_t)_H+\frac{3m_t^4}{4\pi^2 v^2}+{\cal O}(y_t^2m_h^2)
\eea
where  $v\simeq 246\gev$ and $y_t$ is the top-quark Yukawa coupling. 

We can compute the finetuning across the $m_t$ threshold and we find
\beq
{\rm FT}[m_h^2|m_t]\simeq \frac{3m_t^4}{\pi^2 v^2 m_h^2}
\eeq
Inserting $m_t=173\gev$ and $m_h=125\gev$ into this equation one finds ${\rm FT}=0.3$ which is ${\cal O}(1)$ as we expect most finetunings to be across thresholds. This is a low finetuning that is consistent with a Natural theory.

One should note that there would be nothing wrong in principle with the SM if the Higgs boson were much lighter. For example if $m_h$ had been found to be less than $\sim 1\gev$ the finetuning value across the top quark threshold would be  ${\rm FT}\gsim 10^3$. In that case, the theory would have failed  its Wilsonian Naturalness test, which would not have been physically impossible but rather highly improbable. Again, it is to be expected that such a large finetuning did not show up. This is reminiscent of a similar conclusion using a different approach to assess Naturalness -- finite Naturalness~\cite{Farina:2013mla} -- applied specifically to the top quark contribution to the Higgs mass.

The summary of this section is that the SM has low finetunings across matchings of EFTs across its mass thresholds and therefore passes its Naturalness test. There are dozens of non-trivial tests that could have come to a different conclusion. This gives confidence that our primary theory at the present (the SM) does not register as a failure in the Naturalness evaluation with which we plan to asses conjectured theories. This is in contrast to  illogically  charging  the SM with a lethal naturalness problem and then finding new theories that do not. The SM is Natural. Or differently said, the SM does not suffer from Unnaturalness. 

\section{Adding a heavy singlet}
\label{sec:singlet}

One of the simplest ways to extend the SM is to add a real singlet scalar $\sigma$ to the spectrum. One can call this theory SM$+\sigma$ for short. The lagrangian is 
\beq
{\cal L}_{SM+\sigma}={\cal L}_{SM}+\frac{1}{2}(\partial_\mu \sigma)^2-\frac{1}{2}m_\sigma^2 \sigma^2-\frac{\eta_\sigma}{2}H^\dagger H\sigma^2+\frac{\lambda_\sigma}{4} \sigma^4
\eeq
Let us suppose that the mass of the $\sigma$-particle is higher than the  masses of the other particles in the spectrum, and let's also call the effective theory that includes the $\sigma$ particle ${\cal L}_{\sigma+}={\cal L}_{SM+\sigma}$. We shall see that if the mass of the $\sigma$ particle is too high then the matching encounters a finetuning problem, matching the discussion of~\cite{Kaplan:1995uv}.

Given the high mass of the $\sigma$ particle we can integrate it out and are left with a low energy lagrangian ${\cal L}_{\sigma-}$ below the $\sigma$-mass threshold which is the SM lagrangian plus many higher dimensional operators, such as ${\cal O}_6=|H|^6$. After some analysis we can see that no operator in  ${\cal L}_{\sigma-}$ suffers from a finetuning of matching across the $m_\sigma$ threshold except possibly the coefficient $m^2$ of the operator $|H|^2$. In that case the matching is
\beq
\label{eq:sigmamatching}
m^2_{(-)}=m^2_{(+)}-\frac{\eta_\sigma m_\sigma^2}{16\pi^2}\left[ 1-\ln\left( \frac{m_\sigma^2}{\mu^2}\right)\right]
\eeq
where for clarity we have defined 
\beq
m^2_{(\pm)}=m^2~{\rm evaluated~at}~q^2=m_\sigma^2 (1\pm \epsilon),~{\rm where}~\epsilon \ll 1.
\eeq
 In other words $m^2_{(-)}$ is the coefficient of $|H|^2$ in the low-energy effective theory just below the $m_\sigma$ threshold after the $\sigma$-particle has been integrated out, and $m^2_{(+)}$ is the coefficient of $|H|^2$ in the high-energy theory above the $m_\sigma$ threshold that includes the $\sigma$ particle.

The calculation of the finetuning across the threshold yields:
\beq
\label{eq:sigmaFT}
{\rm FT}[m^2]=\left| \frac{m_\sigma^2}{m^2}\frac{\partial m^2}{\partial m_\sigma^2}\right|=
\left| \frac{\eta_\sigma m^2_\sigma}{8\pi^2m_h^2}
\ln\left( \frac{m_\sigma^2}{\mu^2}\right)\right|
\eeq
where the last term utilizes the leading order result in the low-energy theory that $m^2=-m_h^2/2$. 
We come across here the first instance where we think carefully how to deal with the arbitrary matching scale $\mu$.

In the computation of observables the final result cannot depend on the arbitrary scale $\mu$ in the loop integrals. In practice, the result does not depend on $\mu$ at the order of the calculation, and a substantial residual $\mu$ dependence is an indication of the importance of including higher order corrections. On the other hand, the computation of the finetuning through sensitivity to loop effects, such as our example in eq.~\ref{eq:sigmaFT}, depends on $\mu$. Not only that, choosing $\mu=m_\sigma$ in eq.~\ref{eq:sigmaFT} appears to conclude that there is no finetuning independent of the mass $m_\sigma^2$. 

Is not then the strong dependence of finetuning on the arbitrary scale choice $\mu$ of matching an indication of the worthlessness of finetuning? No. The proper way to think of the scale choice is that it is indeed arbitrary, and any reasonable choice from the point of view of theory calculability can be employed and the result should not be finetuned across the boundary. Reasonable choices the community has long made is $m_\sigma/2<\mu<2m_\sigma$, which translates into $\ln (m_\sigma^2/\mu^2)$ ranging from $-1.4$ to $1.4$. Thus, we should vary $\mu$ over the entire range and determine what the average finetuning is. That is one precise prescription on how to deal with $\mu$ dependence. Another prescription which is much simpler to implement and yields roughly the same results is to replace $\ln(\Delta/\mu^2)$ with both $+1$ and $-1$ and average over the two resulting values for finetuning. We go with that prescription unless there is a case where replacing with $+1$ and $-1$ gives artificial cancellations, and then we revert back to the original prescription of averaging finetuning over choices $-1.4<\ln(\Delta/\mu^2)<1.4$. 
Our singlet scalar example here has no artificial cancellations by replacing $\ln(\Delta/\mu^2)=\pm 1$ and this prescription applied to our example then gives
\beq
\label{eq:FTm2}
{\rm FT}[m^2]=
\frac{\eta_\sigma m^2_\sigma}{8\pi^2m_h^2}=0.8\, \eta_\sigma \left( \frac{m_\sigma}{1\tev}\right)^2
\eeq

The reader may be concerned that by using $\ln(\Delta/\mu^2)=\pm 1$ we changed the definition of our finetuning requirement by evaluating $\mu$ away from the precise mass threshold. However, this was only done to maintain the intuitions of those who wish to determine the finetuning of the Higgs boson with respect to $m^2_\sigma$. There is no problem maintaining the strict requirement $\mu=m_\sigma$ and seeing that there is large finetuning. Recall, the finetuning is the maximum finetuning across a threshold with respect to any high-energy parameter. The mass $m_\sigma^2$ was one such  parameter but so is the $H^\dagger H$ operator coefficient $m^2_{(+)}$ in the high-scale theory. The finetuning with respect to it is
\bea
{\rm FT}[m^2]&=&\FTtwo{m^2_{(-)}}{m^2_{(+)}}=\left| \frac{m^2_{(+)}}{m^2_{(-)}}\frac{\partial m^2_{(-)}}{\partial m^2_{(+)}}\right|= \left| \frac{m^2_{(+)}}{m^2_{(-)}}\right|_{\mu^2=m_\sigma^2} =\left| 1-\frac{\eta_\sigma m_\sigma^2}{8\pi^2m_h^2}\right| \nonumber \\
&\simeq & \frac{\eta_\sigma m_\sigma^2}{8\pi^2m_h^2}~~{\rm (for~large~}m_\sigma^2{\rm )}
\eea
which is the same result as eq.~\ref{eq:FTm2}. The reason for this high finetuning seeping into the $m^2_{(+)}$ dependence is that the large logarithm is none other than the $\beta$-function of the renormalization group flow for the $m^2_{(+)}$ mass parameter. Very large changes in $m^2_{(+)}$ over small changes in $\mu^2$ is what ultimately led to the finetuning with respect to $m^2_{(+)}$ evaluated at the $\mu^2=m_\sigma^2$ scale.

Therefore, we can with confidence use eq.~\ref{eq:FTm2} to determine the finetuning across the $m^2_\sigma$ mass threshold. With this in hand, we find that to reach level-3 finetuning or above with $\eta_\sigma=1$ the $\sigma$ mass must be $m_\sigma> 36\tev$. This Naturalness proscription against such high masses might upon first inspection seem remote and uninteresting, but it is a remarkably powerful constraint that in all the many mass scales of nature, from here to the Planck scale, the introduction of any scalar boson that interacts moderately well with the SM Higgs would create, according to our assessments, a theory that does not pass its Wilsonian Naturalness test across the mass threshold. This is another articulation of the serious ``proliferation naturalness problem" that was discussed in sec.~\ref{sec:SM}, and it is why many wish nature to possess a deeper principle, such as supersymmetry, to automatically allow large numbers of additional states, beyond the SM states that we know about, to be present in nature without facing repeated worries about how the Higgs boson mass could be so light in the presence of them all.

\section{Supersymmetry}

The case of supersymmetry adds some complication to the algorithm of computing finetunings across thresholds. In the case of the minimal supersymmetric Standard Model (MSSM) the largest threshold finetuning is going from a full Higgs sector with two Higgs doublets to a low-energy theory with a single Higgs doublet that has the propagating $h125$ boson and the three Goldstone bosons that constitute the longitudinal components of the $W^\pm$ and $Z^0$. Our goal then is to describe the matching between the full MSSM lagrangian parameters with two Higgs doublet ($H_u$, $H_d$) and the effective theory below a heavy lepton multiplet ($\Phi=\{ A,H^0,H^\pm\}$, where the low-energy theory contains the SM doublet $H$.

As is well known, as superpartner masses increase the heavy doublet decouples with it and can be thought increasingly as a vev-less heavy complex scalar. This can then be eliminated. This heavy state can then be integrated out and the low-scale lagrangian that results can match its parameters with those of the theory with the heavy state. The subtlety is what algorithm to use for the ``lagrangian parameters" of the full MSSM theory. In this supersymmetric case the most straightforward choice, and the choice that has been considered for many applications before, are the gauge couplings ($g'$, $g$, and $g_s$), superpotential parameters (Yukawa couplings and the $\mu$ term), and soft supersymmetry breaking terms ($m_{H_u}^2$, $m_{H_d}^2$, $m_{\tilde Q}^2$, etc.)

With these considerations we first write the theory above the heavy Higgs doublet threshold~\cite{Martin:1997ns}:
\bea
V(H_u,H_d)&=& (|\mu|^2+m^2_{H_u})|H_u|^2+(|\mu|^2+m^2_{H_d})|H_d|^2-b H_u\cdot H_d+{\rm c.c.}   \nonumber \\
& & +\frac{1}{8}g'^2\left( |H_u|^2-|H_d|^2\right)^2+\frac{1}{8}g^2\left( H_u^\dagger\sigma^aH_u+H_d^\dagger\sigma^aH_d\right)^2
\eea
This then needs to be matched to the theory below the heavy Higgs doublet threshold
\beq
\label{eq:Vhiggs}
V(H)=m^2|H|^2+\lambda |H|^4
\eeq
After some manipulations one finds
\bea
\label{eq:m2}
m^2 & = & -\left( \frac{1-\sin^2 2\beta}{2}\right)\left[ \frac{|m_{H_d}^2-m^2_{H_u}|}{\sqrt{1-\sin^2 2\beta}}-m_{H_d}^2-m_{H_u}^2-2|\mu|^2\right]   \\
\label{eq:lambda}
\lambda &=& \frac{1}{8}\left( g'^2+g^2\right)(1-\sin^2 2\beta) 
\eea
where
\beq
\label{eq:s2beta}
\sin 2\beta=\frac{2b}{m_{H_d}^2+m_{H_u}^2+2|\mu|^2}.
\eeq
Given this definition of the angle $\beta$, in the limit of large superpartner mass scale we can identify the light SM Higgs boson $H$ and the heavy decoupled doublet state $\Phi$ with
\bea
H & = &  \cos\beta H_d+\sin\beta \overline{H}_u \\
\Phi & = & -\sin\beta H_d +\cos\beta \overline{H}_u.
\eea
where $\overline{H}_u=i\sigma^2 H_u^*$.

We are left with the below-threshold lagrangian of~eq.~\ref{eq:Vhiggs} and its matching equations (eqs.~\ref{eq:m2} and~\ref{eq:lambda}) entirely in terms of the parameters of the above-threshold theory once the dependence of that angle $\beta$ on the supersymmetry parameters are substituted (eq.~\ref{eq:s2beta}). If we identify the heavy supersymmetry masses as the collection of $\tilde m^2_k=\{ m_{H_u}^2,m_{H_d}^2,|\mu|^2,b\}$ that have typical scale values of $\Lambda_{\rm susy}$, one can see readily from eq.~\ref{eq:m2} that
\beq
{\rm FT}[m^2]={\rm max}_k \left| \frac{\tilde m_k^2}{m^2}\frac{\partial m^2}{\partial \tilde m_k^2}\right| \sim \frac{\Lambda^2_{\rm susy}}{m_Z^2}.
\eeq
The precise values of the FT depends on the exact choices of parameters but it is generic that the result is as shown, ${\rm FT}\sim \Lambda_{\rm susy}^2/m_Z^2$.

This scaling of finetuning matches the intuitions that have been present in the supersymmetry community for quite some time now. Traditionally the calculation was to check on the finetuning of the small value of $m_Z^2$ given all the heavy superpartner masses in the scalar potential. The equation for $m_Z^2$ for electroweak symmetry breaking at leading order is
\beq
m_Z^2=-2|\mu|^2+\frac{2(m_{H_d}^2-m_{H_u}^2\tan^2\beta)}{\tan^2\beta-1}
\eeq
Finetunings are then computed and the result is generically ${\rm FT}\sim \Lambda_{\rm susy}^2/m_Z^2$. So, although the ${\cal O}(1)$ factors will be different between our algorithm for computing the threshold finetuning and the finetuning computed from considering superpartner mass dependences on $m_Z$, the results are the same within ${\cal O}(1)$ factors. The main reason for this is that at tree-level $m^2=-\frac{1}{2}m_Z^2\cos^22\beta$, and so computations of finetuning on $\tilde m^2$ should be very similar to that of $m_Z^2$. 

Roughly speaking using the guide that ${\rm FT}\sim \Lambda_{\rm susy}^2/m_Z^2$ one can contemplate declaring some supersymmetric theories\footnote{Remember, a theory in our definition here does not have variable parameters. Theories are indexed values of a theory class. A supersymmetric theory then has specific single values for each of its parameters $m^2_{H_u}$, $\mu$, etc.\ defined at some convenient scale.}
with large $\Lambda_{\rm susy}$ to fail the Naturalness test and thus should be relegated to lower status. Less probable, and perhaps even improbable, level-4 finetuning, for example, would put $\Lambda_{\rm susy}^2/m_Z^2\sim 10^4$ and thus $\Lambda_{\rm susy}\sim 10\tev$. LHC limits of $1-3\tev$ for the mass of superpartners puts the current level of finetuning at about $10^{2}-10^{3}$, which is borderline for when someone would wish to confidently relegate the remaining supersymmetric theories to the trash bin. Note, there still remains an infinite number of supersymmetric theories that perfectly satisfy all known data and are not in conflict with any expectations. Only a high level of finetuning across thresholds (i.e., failing its Wilsonian Naturalness test) can cast a shadow on any of those otherwise good theories. It is with trepidation that one throws out otherwise perfectly good theories that are consistent with all known data; nevertheless, we reiterate adherence to the notion that such high finetunings are correlated with low {\it a priori} probabilities for realization~\cite{Wells:2018yyb}. 

In concluding this section, let us briefly remark that our approach differs significantly from some others in assessing finetuning in supersymmetric theories in that here we only assess finetunings among parameters ``locally" across a threshold. We look at parameters above and below a heavy mass threshold and ask if there is finetuning in that matching. Some have implemented ``non-local" finetuning assessments, by considering a supersymmetric theory with input parameters, say $m_{1/2}$ (universal gaugino mass) and others~\cite{Bastero-Gil:1999jqv}, defined at the GUT scale and then tracking how $m_{H_u}^2$ and $m_{H_d}^2$ scale with it: $m_{H_{u,d}}^2=m_{H_{u,d}}^2(m_{1/2})$. Finetunings are then computed with respect to the input parameter $m_{1/2}$:
\beq
{\rm FT}[m_Z^2]=\left| \frac{m_{1/2}}{m_Z^2}\frac{\partial m_Z^2}{\partial m_{1/2}}\right|.
\eeq
In this case there is implicitly a calculation of a parameter at one scale ($m_Z$ at $q^2=m_Z^2$) in terms of a parameter $m_{1/2}$ defined far away at the scale $q^2=M^2_{\rm GUT}$.
Related schemes for identifying independent vs.\ dependent variables with correlations at various scale choices among the supersymmetric parameter choices may be introduced, which in our language would be useful for turning a theory that fails to be endo-Natural into one that is exo-Natural, and thus respects Wilsonian Naturalness. We are neutral as to the value of this activity, although we recognize that conceivably it could be used to both artificially and inappropriately lower the finetuning of a theory, but also to identify underlying parameter correlations that yield lower finetuning, which a deeper theory might be able to justify. With that said, the intuition at play here that requires a speculative theory to be Wilsonian Natural is laudable and justified.

\section{Grand Unified Theories}

Another example of an improbability of parameter cancellations that has been discussed in the literature for years is the so-called doublet-triplet splitting problem in grand unified theories~\cite{Mohapatra:1997sp,Raby:2017ucc}. For example, minimal $SU(5)$ theory breaks down to the SM gauge groups via the condensation of the 24 dimensional representation $\Sigma$. The vacuum expectation value of this field is
\beq
\langle\Sigma\rangle=v_\Sigma\cdot{\rm diag}(2,2,2,-3,3)
\eeq
where the value of the vev $v_\Sigma$ is determined by parameters $\vec w$ in GUT-scale Higgs potential: $v_\Sigma=v_\Sigma(\vec w)$.

In the supersymmetric case the $\Sigma$ also couples to the $5$- and $\bar 5$-dimensional Higgs representation $H_5$  and $H_{\bar 5}$ respectively. Within the $H_{5,\bar 5}$ are the Higgs doublets $H_{u,d}$ and the Higgs triplet $H_{3,\bar 3}$ representations. The relevant GUT-scale superpotential for $H_5$ is
\beq
W_{(+)}=\mu_5 H_{\bar 5} H_5+\lambda H_{\bar 5} \Sigma H_5
\eeq
After symmetry breaking the superpotential splits the $H_{5,\bar 5}$ into $H+{u,d,3,\bar 3}$ terms:
\beq
W=\mu_3 H_{\bar 3}H_3+\mu H_uH_d\Longrightarrow W_{(-)}=\mu H_uH_d+\cdots
\eeq
where
\bea
\mu_3& = & \mu_5+2\lambda v_\Sigma,~~{\rm and} \\
\mu& = &\mu_5-3\lambda v_\Sigma. \label{eq:mu}
\eea
We know that $v_\Sigma\simeq 10^{16}\gev$ for the unification of couples, and we also know that $\mu$ needs to be $10^{2-3}\gev$ for weak scale supersymmetry. Thus, there is an extraordinary finetuning in the cancellation that must occur in eq.~\ref{eq:mu} to realize these constraints. Upon symmetry breaking and assessing the finetuning of $\mu$ with respect to the high-scale theory parameter $\mu_5$ one finds
\beq
{\rm FT}[\mu]=\left| \frac{\mu_5}{\mu}\frac{\partial \mu}{\partial \mu_5}\right|= \left|\frac{\mu_5}{\mu}\right|\simeq \left| \frac{2\lambda v_\Sigma}{\mu}\right|\sim 10^{13}
\eeq
Thus,  minimal supersymmetric $SU(5)$ GUTs have level-13 finetuning and do not pass their Wilsonian Naturalness test. This is why it is often referred to in the literature as the doublet-triplet splitting problem. It really is simply a Wilsonian Naturalness problem of the theory across matching EFT thresholds.

Of course there are many interesting ideas on how to solve this problem in Grand Unified theories, but the point here is that it is identified as a problem immediately from the perspective of our algorithmic finetuning tests for Wilsonian Naturalness, and the magnitude of the problem (level-13 finetuning) matches intuitions of early researchers in GUT theories who appreciated the seriousness of this deficiency in minimal GUT theories.

\section{Extra dimensions and the hierarchy problem}
\label{sec:Xdim}

Let us make a few brief remarks about theories of extra dimensions~\cite{ArkaniHamed:1998rs,Antoniadis:1998ig,Randall:1999ee}. Researchers have added extra spatial dimensions, either flat or warped, in order to recast and perhaps solve the hierarchy problem. It is worthwhile making a rather precise definition of the hierarchy problem here in order to give our perspective on the worth of introducing extra spatial dimensions to solve it.

If the SM does not have a Wilsonian Naturalness problem, as we argued sec.~\ref{sec:SM}, then what can we mean that it has a hierarchy problem, especially since the terms are often used interchangeably in the literature? The definition proposed here is that a theory has a hierarchy problem if it fails to pass its Naturalness tests (low finetunings across thresholds) {\it or} the simple introduction of heavy states into the spectrum immediately creates a Naturalness problem for the theory (``the proliferation of states problem" of  ref.~\cite{Wells:2016luz}). The SM passes its Wilsonian Naturalness test, but if we add an additional scalar with mass $100\tev$, our results from sec.~\ref{sec:singlet} tell us that this new theory will fail its Wilsonian Naturalness test. In that sense the SM has a hierarchy problem -- its passing of the Wilsonian Naturalness test is precarious in the face of nature having additional heavy particles well above the weak scale, which we generically expect nature to possess.

The introduction of supersymmetry solves the hierarchy problem by introducing superpartners that cancel out the quadratic sensitivities of new heavy mass scales to the Higgs parameter. Composite Higgs theories solve the hierarchy problem by eliminating this most special quadratically sensitive parameter associated with a fundamental Higgs boson mass~\cite{Panico:2015jxa}. And now, here, extra dimensions also could solve the hierarchy problem by disallowing any state to have mass that is well above the weak scale. There is no singlet scalar mass of $45\, {\rm PeV}$ that would destabilize the hierarchy because the highest, fundamental scale $\Lambda_F$ accessible to field theory is the weak scale, by virtue of, for example, an exponential suppression $\Lambda_F=M_{\rm Pl}e^{-y}$, where $y\simeq 34$ associated with a compactified warped extra dimension. 

Therefore, large extra dimension or compactified extra dimensions do not cure the SM of its Naturalness problem, because the SM has no Naturalness problem, but it does replace our Standard Theory of the SM with a different theory of extra dimensions, which, unlike the SM, has no hierarchy problem if the scale of extra dimensions is small enough. Extra dimensional theories have little to speak for themselves otherwise, so their advocacy has implicitly assumed in it that nature should have other states besides the SM states, and when those other states are added to the SM there develops a significant finetuning problem across its EFT matching thresholds, and thus the SM will not pass its Wilsonian Naturalness test without the introduction of extra spatial dimensions to cut off the offending high scales.

\section{Twin Higgs theories}

Another class of theories that is reported to stabilize the weak scale is Twin Higgs theories~\cite{Chacko:2005pe,Low:2015nqa,Barbieri:2016zxn}. The idea is to introduce a global $U(4)$ in the Higgs sector that the $\Phi$ scalar transforms under. 
The original potential for $\Phi$ admits spontaneous symmetry breaking,
\beq
V(\Phi)=-m^2_\Phi|\Phi|^2+\lambda_\Phi|\Phi|^4~~{\rm where}~~\langle\Phi\rangle=\frac{m^2_\Phi}{2\lambda_\Phi}=f^2,
\eeq
which breaks $U(4)\to U(3)$, giving 7 Goldstone bosons in the process.

Now, assume that the $SU(2)_A\times SU(2)_B$ subgroup of $U(4)$ is gauged and the $\Phi$ field can be split into
\beq
\Phi=\left( \begin{array}{c} H_A \\ H_B\end{array}\right)
\eeq
where $SU(2)_A$ is ultimately identified with the SM $SU(2)_L$ gauge symmetries and $SU(2)_B$ is a ``twin $SU(2)$". $H_A$ then will become the SM Higgs boson and $H_B$ the ``twin partner boson."  One also assumes a $Z_2$ ``twin symmetry" that enforces invariance under exchange of $H_A\leftrightarrow H_B$. This requirement implies equivalence of the $SU(2)_A\times SU(2)_B$ gauge couplings, $g_A=g_B=g$.

Radiative corrections from gauge fields to the potential of this theory will yield terms proportional to 
\bea
\Delta V_2(H_A,H_B)& \sim & \frac{g^2\Lambda^2}{16\pi^2}(|H_A|^2+|H_B|^2)=\frac{g^2\Lambda^2}{16\pi^2}\Phi^\dagger\Phi~~{\rm and}\\
\Delta V_4(H_A,H_B) &= &\frac{g^4}{16\pi^2}\ln\left( \frac{\Lambda^2}{f^2}\right)(|H_A|^4+|H_B|^4).
\eea
where $\Lambda\sim 4\pi f$.
The first terms $\Delta V_2$ do not violate $U(4)$ symmetry but the $\Delta V_4$ terms do, turning the Goldstone bosons into pseudo-Nambu-Goldstones (pNGB). The estimate of the physical Higgs mass is then
\beq
\label{eq:twinmh}
m_h^2\sim \frac{g^4}{16\pi^2}f^2\sim\left( \frac{g^2}{16\pi^2}\right)^2\Lambda^2.
\eeq
Therefore, even for very large $\Lambda$ of several TeV -- high-mass ``new physics" -- the Higgs is two-loop suppressed compared to the scale $\Lambda$ and stays light.

The problem with the scenario above is that the there is complete alignment of the vevs for $H_A$ and $H_B$, where both have $\langle H_{A,B}\rangle\sim f$, which is too high. Also, the SM Higgs, which is a pNGB of the breaking, is an equal admixture of the $SU(2)_A=SU(2)_L$ charged state and the $SU(2)_B$ charged state, and so its couplings to the SM states will be greatly reduced compared to the SM expectations. This is unacceptable given current constraints on the discovered $h125$ state at CERN which shows that the couplings of $h125$ to SM states are at least within $10\%$ of SM couplings.

The solution is to introduce a soft breaking potential
\beq
V_{\rm soft}(\Phi)=\mu^2_\Phi \Phi^\dagger \Phi
\eeq
which explicitly breaks the $Z_2$ twin symmetry of $H_A\leftrightarrow H_B$. It is called ``soft" because this term is technically natural in that when $\mu_\Phi\to 0$ one recovers a higher symmetry (the $Z_2$ twin symmetry). The effect of this is to misalign the vevs such that $\langle H_A\rangle \ll \langle H_B\rangle$, and the light Higgs becomes much more SM-like.

Now, what about the naturalness qualities of this theory? One can readily see by the discussion above that $f$ can be in the multi-TeV range and this Twin Higgs theory would still passes its Wilsonian Naturalness test of computing the finetuning of the low-scale theory parameter $m_h$ (coefficient of $|H|^2$ operator) with respect to the high-scale theory parameters $m^2_\Phi$, $\mu_\Phi$, etc. Thus, the twin Higgs theory does not have a Naturalness problem {\it as long as} the theories mass parameters are not above tens of TeV.

But what has the Twin Higgs theory done for us, vis-\`a-vis naturalness and finetuning? From the algorithmic Wilsonian Naturalness test introduced here in this work the Twin Higgs theory has no impact. It does not solve the SM's Naturalness problem, because the SM has no Naturalness problem to begin with. And it does not solve the Hierarchy problem because if one adds another singlet $\sigma$ to the spectrum and attaches it to $|\Phi|^2\sigma^2$ then a heavy $m_\sigma$ will cause the Twin Higgs$+\sigma$ theory to fail its Wilsonian Naturalness test  at the same disastrous level as SM$+\sigma$ theory does. 

What the Twin Higgs theory does do is soften the quadratic divergences from a cutoff regulation scale from top quark and gauge boson loops. Instead of the Higgs boson being sensitive to $m_h^2\sim (y_t^2/16\pi^2)\Lambda^2$ it is softened to $m_h^2\sim (g^2/16\pi^2)^2\Lambda^2$, enabling $\Lambda$ to be much higher before the destabilizing finetuning effects of a large $\Lambda$ manifest themselves. For those who believe that there is meaning in tracking cutoff scale dependences in an effective theory, and taking them as serious indications of a destabilizing impact of quantum corrections {\it within the theory itself}, the Twin Higgs theory delays the requirement of new physics (UV completion) that will take care of the issue once and for all, such as supersymmetry, conformal symmetries or extra dimensional theories could do. For those who do not put stock in gaining intuitions by naive applications of cutoff dependent regulated quantum corrections, the Twin Higgs theory is merely another interesting theory that has little to do with Naturalness. 

\section{Conclusions}

In this article we have presented a straightforward methodology for testing if a theory is Natural based on finetuning assessments across effective theories above and below massive particle thresholds. The resulting Wilsonian Naturalness test satisfies the requirement of being an {\it a priori} and unambiguous algorithm, which is required for it to have a connection to probability, as argued in~\cite{Wells:2018yyb}. 

In the process we have articulated the difference between a theory that has a Naturalness problem and one that has a Hierarchy problem. A theory has a Naturalness problem if it fails to pass its tests of low finetunings across EFT thresholds. A theory has a Hierarchy problem if it immediately develops a Naturalness problem in the presence of very massive additional particles added to the spectrum, such as a massive real scalar or a vectorlike fermions.

This approach unifies the understanding of traditionally claimed Naturalness problems based on what before appeared to be different criteria, such as within supersymmetry (finetuning of $m_Z$), the SM theory augmented by a massive real scalar (finetuning of $m_h$), and grand unified theories (finetuning of $\mu$-term in GUT superpotential). All three of these Naturalness concerns have straightforward intepretations and are unavoidably accounted for within Wilsonian Naturalness. 

Another implication of the Wilsonian Naturalness test described here is the unambiguous conclusion that the Standard Model does not suffer from a Naturalness problem, in contrast to many statements in the literature. Furthermore, within the  Wilsonian Naturalness approach the twin Higgs theories neither solves the SM's Naturalness problem, because it does not have one, nor does it solve the Hierarchy problem since twin Higgs theories are just as susceptible to failing its Wilsonian Naturalness test in the presence, for example, of a PeV real scalar as is the SM. Nevertheless, the claim here is not that the Wilsonian Naturalness  test is necessarily the only possible Naturalness assessment, but the rigor of other assessments that claim a theory is not Natural must explain its {\it a priori} algorithms and its connection to probabilities, which no other assessments do. 

In summary, Wilsonian Naturalness is firmly grounded in its algorithmic {\it a priori} formulation, and it is connected to probability tests (failing strict Naturalness test is low probability). Therefore, we believe the ``next good theory" beyond the Standard Model whose peculiar features will be corroborated by experiment will very likely be Wilsonian Natural, which in turn can be a guide to whether a new theory competing for attention should have high status among all the theories compatible with known data.

\bigskip

\noindent
{\bf Acknowledgements.}
This work is supported in part by DOE grant DE-SC0007859. I wish to thank S.~Martin and Z.~Zhang for enlightening conversations on these topics.


\end{document}